\documentstyle[twoside,11pt,epsfig,color]{article}
\begin{document}

\title{Gravitational collapse in f(R) gravity for a spherically symmetric spacetime admitting a homothetic Killing vector}

\author{Soumya Chakrabarti\footnote{email: adhpagla@iiserkol.ac.in}~~~and~~Narayan Banerjee\footnote{email: narayan@iiserkol.ac.in}
 \\
Department of Physical Sciences, \\Indian Institute of Science Education and Research, Kolkata\\ Mohanpur Campus, West Bengal 741252, India.
\date{}
}

\maketitle
\vspace{0.5cm}
{\em PACS Nos. 04.50.Kd; 04.70.Bw
\par Keywords : gravitational collapse, $f(R)$ gravity, spherical symmetry, self-similar}
\vspace{0.5cm}

\pagestyle{myheadings}
\newcommand{\be}{\begin{equation}}
\newcommand{\ee}{\end{equation}}
\newcommand{\bea}{\begin{eqnarray}}
\newcommand{\eea}{\end{eqnarray}}

\begin{abstract}
The gravitational collapse of a spherical distribution, in a class of f(R) theories of gravity, where f(R) is power function of R, is discussed. The spacetime is assumed to admit a homothetic Killing vector. In the collapsing modes, some of the situations indeed hit a singularity, but they are all covered with an apparent horizon. Some peculiar cases are observed where the collapsing body settles to a constant radius at a given value of the radial coordinate. 
\end{abstract}

\section{Introduction}

In spite of its great success as a theory of gravity, general relativity (GR) has some alternative theories which come from various motivations. A most recent motivation certainly is the quest for a theoretical framework which can explain the present accelerated expansion of the universe without invoking any exotic matter field. The present interest in $f(R)$ gravity also stems from the same motivation. An $f(R)$ theory is one where the Ricci scalar $R$ in the Einstein-Hilbert action is replaced by an analytic function of $R$. For a comprehensive review of the motivation, various forms and applications of $f(R)$ theories, we refer to the work of Sotiriou and Faraoni\cite{faraoni}. As a theory of gravity, $f(R)$ theories warrant elaborate investigation in the context of local astronomical implications as well. In fact Clifton and Barrow\cite{john} showed that for an $f(R)$ theory, the departure from general relativity has to be very small so as to be compatible with local astronomy. Also quite often it is difficult to find a Schwarzschild analogue\cite{dolgov} as a stable solution. However, recently Nzioki, Goswami and Dunsby showed that this latter problem is not generic and a Schwarzschild solution is in fact a stable limit of a section of $f(R)$  theories\cite{rituparno1}. \\

The motivation of the present work is to look at the collapsing scenario in $f(R)$ gravity in a spherically symmetric spacetime. Gravitational collapse, by its own right, forms an important arena of research. A stellar object, after it burns up its nuclear fuel, may not have sufficient pressure so as to balance the gravitational pull of its own mass and collapses to a tiny volume. The subject of interest is to check whether it collapses to a singularity, and if it does whether the singularity is hidden from an observer faraway by means of a horizon. In general relativity, the theoretical study of a collapsing stellar object started with the work of Oppenheimer and Snyder\cite{opp}, where the unhindered collapse of a perfect fluid with no pressure was discussed (see also Datt\cite{datt}). The importance and various implications of gravitational collapse is reviewed recently by Joshi. The technical aspects of collapse, particularly with a fluid, can be found in the monograph by Joshi\cite{pankaj}. \\

Although various aspects of $f(R)$ gravity theories have been studied quite extensively, interest in collapsing models in such theories came into being quite recently. Rigorous analytical investigation in this direction has been carried out very recently by Goswami {\it et al}\cite{rituparno2}. This brings out some important results, for instance the existence of a regular horizon would require an inhomogeneity in the fluid distribution. Very recently Chakrabarti and Banerjee\cite{soumya} investigated an $f(R)$ gravity collapse in a spherically symmetric case where the metric coeffiecients were assumed to be separable function of the radial coordinate $r$ and the time coordinate $t$; the field equations yielded some restrictive domain of the validity as $\frac{df(R)}{dR}$ came out to be separable as well. The requirement of inhomogeneity is strongly supported in the latter work as well. The stability of collapsing models in $f(R)$ theories has been investigated by Sharif and Yousaf\cite{sharif1, sharif2}. Apart from these analytical work, numerical techniques have also been utilized in $f(R)$ gravity collapse by Borisov, Jain and Zhang\cite{borisov} and also by Guo, Wang and Frolov\cite{guo}. \\

In the present work we start with a general form of $f(R)$, but eventually make an attempt to look for solutions for $f(R) \sim R^{n}$, for as general value of $n$ as possible. The equation system, obviously, is too difficult to handle analytically in general. We impose certain symmetry condition on the spacetime metric so as to arrive at some conclusions analytically. We impose the existence of a homothetic Killing vector at the outset. This will definitely restrict the geometry, but keep alive the dependence on the radial coordinate $r$ which is indeed very much crucial in keeping the option open for a formation of horizon in the case of an ultimate singularity as a result of the collapse. The existence of a homothetic Killing vector results in a simplification of the metric which has nicely been summarized and utilized by Wagh and Govinder\cite{wagh}.

In section 2 we give a very brief description of a spherically symmetric spacetime admitting a homothetic Killing vector. Section 3 describes the formalism of a metric $f(R)$ gravity. Section 4 discusses the exact analytic collapsing solutions when $f(R)$ is a power function of the Ricci scalar $R$ in terms of Gauss' hypergeometric functions. The 5th section discusses the formation of singularity, the nature of the singularity and the possibility of the formation of an apparent horizon. The 6th section deals with the matching of the solutions obtained with an stable exterior metric. The last section summarizes the results obtained.

\section{A Homothetic Spacetime}

A vector field $X$ is called a Killing vector field if it satisfies the Killing equation given as:
\begin{equation}\label{killing}
L_{X}g_{ab}=g_{{ab},c}X^c+g_{cb}X^c,a+g_{ac}X^c,b=0,
\end{equation}
where $L$ is the Lie derivative operator along the vector field $X$. The existence of a Killing vector signifies a symmetry and hence a conserved quantity. For example, if a timelike Killing vector exists for a particular metric, the energy is conserved. \\

If the right hand side of equation (\ref{killing}) is not zero but proportinal to $g_{ab}$, 
\begin{equation}\label{homothety}
L_{x}g_{ab}=2\Phi g_{ab},
\end{equation}
$X$ is called conformal Killing vector. Here $\Phi$ is a function, called the conformal factor. The space time admits a conformal symmetry in this case. If $\Phi$ is a constant, the vector $X$ is called homothetic Killing vector. A spacetime admitting a homothetic Killing vector is called self-similar, where one can have repetitive structures at various scales. With a conformal symmetry the angle between two curves remains the same and only the distance between two points are scaled by a factor depending on the spacetime points. For self-similarity, the scaling of the distance is a constant. \\

A spherically symmetric self-similar spacetime can be shown to admit a homothetic Killing vector of the form

\begin{equation}\label{homothety2}
X^a=(0,f(r,t),0,0).
\end{equation}

The choice of this form is not the general practice, which is rather like a choice as $X^a = (t,r,0,0)$ in a suitable coordinate system. One can show that with the choice of the Homothetic Killing vector as in (\ref{homothety2}), the metric components are all separable functions of the radial coordinate $r$ and the time coordinate $t$, and the metric is given by 

\begin{equation}\label{metric}
ds^2=y^2(r)dt^2-2B^2(t)\Bigg(\frac{dy}{dr}\Bigg)^2dr^2-y^2(r)B^2(t)d\Omega^2,
\end{equation}

and it admits a spacelike homothetic killing vector of the form $\Bigg(0, \frac{y}{\sqrt{2}y'},0,0\Bigg))$. For a comprehensive discussion on this, we refer to the work of Wagh and Govinder\cite{wagh}. \\

Ricci Scalar for the metric (\ref{metric}) can be calculated to be
\begin{equation}\label{ricci}
R=\frac{1}{y^2(r)}\Bigg(\frac{1}{B^2}-6\frac{\dot{B}^2}{B^2}-6\frac{\ddot{B}}{B}\Bigg),
\end{equation}
where an overhead dot and a prime indicate differentiation with respect to time $t$ and the radial coordinate $r$ respectively. It is interesting to note that the Ricci scalar also turns out to be separable in functions of $t$ and $r$ for this class of spacetime line elements.

\section{Formalism of f(R) Gravity}
In $f(R)$ theories, the Einstein-Hilbert action of General Relativity is modified by using a general analytic function $f(R)$ instead of $R$. The action is given by

\begin{equation}\label{action}
A=\int\Bigg(\frac{f(R)}{16\pi G}+L_{m}\Bigg)\sqrt{-g}~d^{4}x,
\end{equation}

where $L_{m}$ is the Lagrangian for the matter distribution.

We work with the standard metric formulation where the action is varied with respect to $g_{\mu\nu}$. The variation of the action (\ref{action}) with respect to the metric tensor leads to the following partial differential equations as the field equations,

\begin{equation}\label{fe}
F(R)R_{\mu\nu}-\frac{1}{2}f(R)g_{\mu\nu}-\nabla_{\mu}\nabla_{\nu}F(R)+g_{\mu\nu}\Box{F(R)}=-8\pi G T^{m}_{\mu\nu},
\end{equation}
where $F(R)=\frac{df}{dR}$.
Writing this equation in the form of Einstein tensor, one obtains
\begin{equation}
G_{\mu\nu}=\frac{\kappa}{F}(T^{m}_{\mu\nu}+T^{C}_{\mu\nu}),
\end{equation}
where $T^{m}_{\mu\nu}$ is the stress energy tensor for the matter distribution defined by $L_{m}$ and 
\begin{equation}\label{curvstresstensor}
T^{C}_{\mu\nu}=\frac{1}{\kappa}\Bigg(\frac{f(R)-RF(R)}{2}g_{\mu\nu}+\nabla_{\mu}\nabla_{\nu}F(R)-g_{\mu\nu}\Box{F(R)}\Bigg).
\end{equation}    \\

$T^{C}_{\mu\nu}$ represents the contribution of the curvature and may formally be treated as an effective stress-energy tensor with a purely geometrical origin. In the present work, we take the stress-energy tensor for the matter part to be that of a perfect fluid which is given by $T^{m}_{\mu\nu}=(\rho+p)v_{\mu}v_{\nu}-pg_{\mu\nu}$. Here $\rho$ and $p$ are the density and pressure of the fluid respectively and $v^{\mu}$ is the velocity four-vector of the fluid particles, which, being a timelike vector, can be normalized as $v^{\mu}v_{\mu} = 1$.  \\

This theory is essentially a nonminimally coupled theory, the curvature related term $F(R)$ couples with the matter sector $T^{m}_{\mu\nu}$ nonminimally.

\section{Exact Solution for $f(R)=\frac{R^{(n+1)}}{(n+1)}$}
From (\ref{metric}) and (\ref{fe}), the $G_{01}$ equation yields
\begin{equation}\label{g01}
2\frac{\dot{B}}{B}\frac{y'}{y}=\frac{\dot{F}'}{F}-\frac{y'}{y}\frac{\dot{F}}{F}-\frac{\dot{B}}{B}\frac{F'}{F}.
\end{equation}

From equation (\ref{ricci}), the derivative of the Ricci Scalar with respect to $r$ can be written in the form
\begin{equation}\label{trick}
R'=-2\frac{y'}{y}R.
\end{equation}

Equations (\ref{g01}) and (\ref{trick}) can be combined to yield
\begin{equation}\label{trick2}
2\frac{\dot{B}}{B}=\Bigg[\frac{2R\frac{d^2F}{dR^2}+3\frac{dF}{dR}}{2R\frac{dF}{dR}-F}\Bigg]\dot{R}=\Lambda(R)\dot{R},
\end{equation}
where $\Lambda(R)$ depends solely on the choice of $f(R)$ (therefore on $\frac{df(R)}{dR}$). \\

In what follows, we choose a particular form of $f(R)$, namely, $f(R)=\frac{R^(n+1)}{(n+1)}$ ($n \neq -1$), so that $F(R)=R^n$. With this choice, equatuin (\ref{trick2}) simplifies to the form

\begin{equation}
\label{trick3}
\frac{\dot{B}}{B}=\frac{n(2n+1)}{2(2n-1)}\frac{\dot{R}}{R}=m\frac{\dot{R}}{R}.
\end{equation}
Here  $m=\frac{n(2n+1)}{2(2n-1)}$.
Hence, it is straightforward to write
\begin{equation}\label{BandR}
B=\delta(r)R^m,
\end{equation}
where $\delta(r)$ is an arbitray function of $r$ arising out of the integration over time.       \\
From the metric (\ref{metric}), one can now write the $G^0_0$, $G^1_1$ and $G^2_2$ equations as

\begin{equation}\label{density}
3\frac{\dot{B}^2}{B^2y^2}=\kappa\frac{\rho}{F}+\frac{f-RF}{2F}+\frac{F''}{2B^2y'^2F}+\frac{\dot{B}\dot{F}}{BFy^2}+\Bigg(\frac{y'}{y}-\frac{y''}{2y}\Bigg)\frac{F'}{FB^2y'^2},
\end{equation}
\begin{equation}\label{press1}
G^1_1=\frac{1}{y^2}\Bigg[2\frac{\ddot{B}}{B}+\frac{\dot{B}^2}{B^2}\Bigg]-\frac{1}{2y^2B^2}=-\kappa\frac{p}{F}+\frac{f-RF}{2F}-\frac{\ddot{F}}{y^2F}+2\frac{\dot{B}\dot{F}}{y^2BF}+\frac{3F'}{2yy'B^2F},
\end{equation}
\begin{equation}\label{press2}
G^2_2=\frac{\dot{B}^2}{y^2B^2}-\frac{1}{2y^2B^2}=-\kappa\frac{p}{F}+\frac{f-RF}{2F}-\frac{\ddot{F}}{y^2F}+\frac{F''}{2B^2y'^2F}+\frac{F'}{2B^2y'^2F}\Bigg(2\frac{y'}{y}-\frac{y''}{y'}\Bigg)
\end{equation}
respectively. The isotropy of the fluid pressure now yields (from equations (\ref{press1}) and (\ref{press2}))
 
\begin{equation}\label{isopress}
2\frac{\ddot{B}}{y^2B}=2\frac{\dot{B}\dot{F}}{y^2BF}+\frac{3F'}{2yy'B^2F}-\frac{F''}{2B^2y'^2F}-\frac{F'}{2B^2y'^2F}\Bigg(2\frac{y'}{y}-\frac{y''}{y'}\Bigg).
\end{equation}       \\

Using equation (\ref{trick}) and (\ref{BandR}) in (\ref{isopress}), one arrives at a non-linear second order differential equation for $B$ as
\begin{equation}\label{timeevolution}
2\ddot{B}B-4\frac{(2n-1)}{(2n+1)}\dot{B}^2+2n(n+1)=0,
\end{equation}

where $\alpha=4\frac{(2n-1)}{(2n+1)}.$ This equation is readily integrated to yield a first integral as

\begin{equation}\label{1stint}
\dot{B}^2=\lambda B^{\alpha}+\frac{2n(n+1)}{\alpha},
\end{equation}
where $\lambda$ is a constant of integration and sensitive to initial geometry profile.    \\

The general solution of (\ref{1stint}) can be written in terms of Gauss' Hypergeometric Function,
\begin{equation}\label{exactsol}
\frac{B}{\sqrt{\frac{2n(n+1)}{\alpha\lambda}}}{_2}F_{1}\Bigg[\frac{1}{2},\frac{1}{\alpha};\Big(1+\frac{1}{\alpha}\Big);-\frac{\alpha\lambda B^\alpha}{2n(n+1)}\Bigg]=\sqrt{\lambda}(t_0-t).
\end{equation}
For a real solution $\Big(1+\frac{1}{\alpha}\Big)>0$, which obviously imposes some restriction over the choice of $n$, i.e. the choice of $f(R)$.  One can note that for $n=0$, one goes back to Einstein gravity, for $n=\frac{1}{2}$, equation (\ref{trick3}) is not valid and for $n=-\frac{1}{2}$, $B$ is constant so there is no evolution. To have a real time-evolution one must impose that $n\notin [-1,-\frac{1}{2}]\cup[0,1).$  \\

\section{Analysis of the solution}
Generally, it is not easy to invert (\ref{exactsol}) to write $B$ explicitly as a function of $t$. So in what follows we try to look at the collapsing modes with the help of numerical plots. Figures $(1)$ and $(2)$ show the behaviour of the collapse for some negative vales of $n$ with $n<-1$ for positive and negative values of the constant $\lambda$. We have the plots for $n=-2$ and $n=-3$. In both the cases the scale factor $B$ and hence the volume decreases with time, but the rate of collapse slows down, and the sphere asymptotically steadies down to a minimum non-zero volume for positive values of $\lambda$. However, for a negative $\lambda$, the radius of the two sphere goes to zero quite rapidly and reaches zero at a finite future. The figures also show that the nature of collapse hardly depends on the particular value of $n$, only the time of reaching the singularity changes. For other negative values of $n$, allowed by the model, plots of exactly similar nature are obtained.      \\

\begin{figure}[t]
\begin{center}
\includegraphics[width=0.4\textwidth]{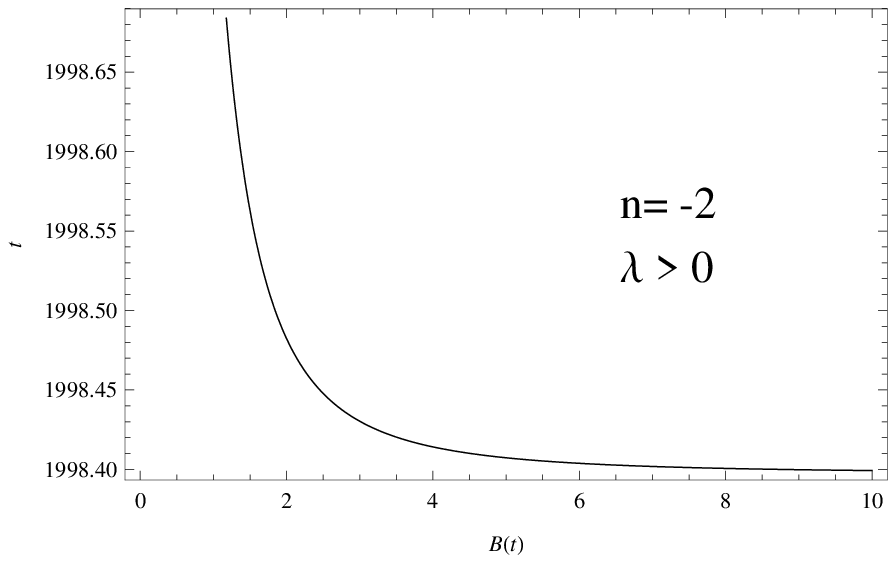}
\includegraphics[width=0.4\textwidth]{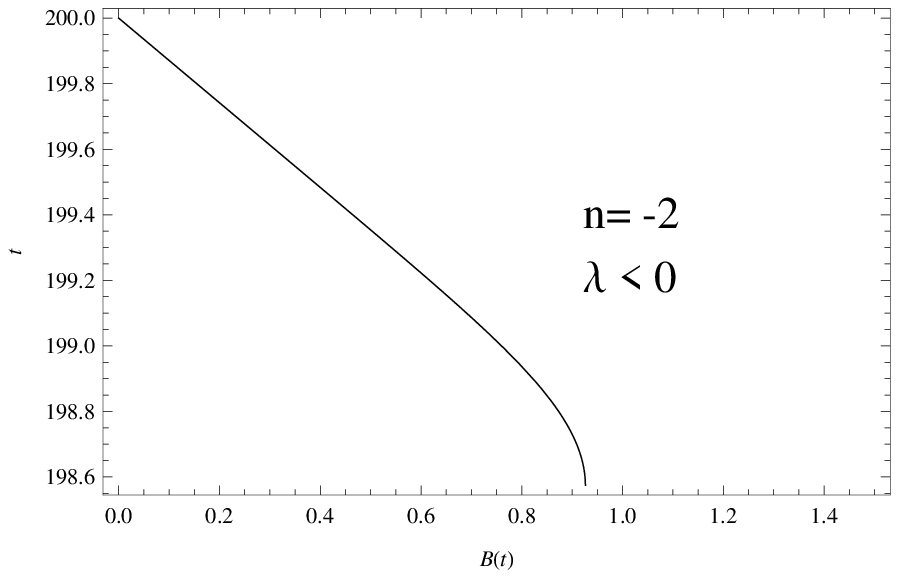}
\caption{Time evolution for different signatures of initial condition ($\lambda$) and $f(R)=-\frac{1}{R}$.}
\end{center}
\label{fig:ss12}
\end{figure}

\begin{figure}[h]
\begin{center}
\includegraphics[width=0.4\textwidth]{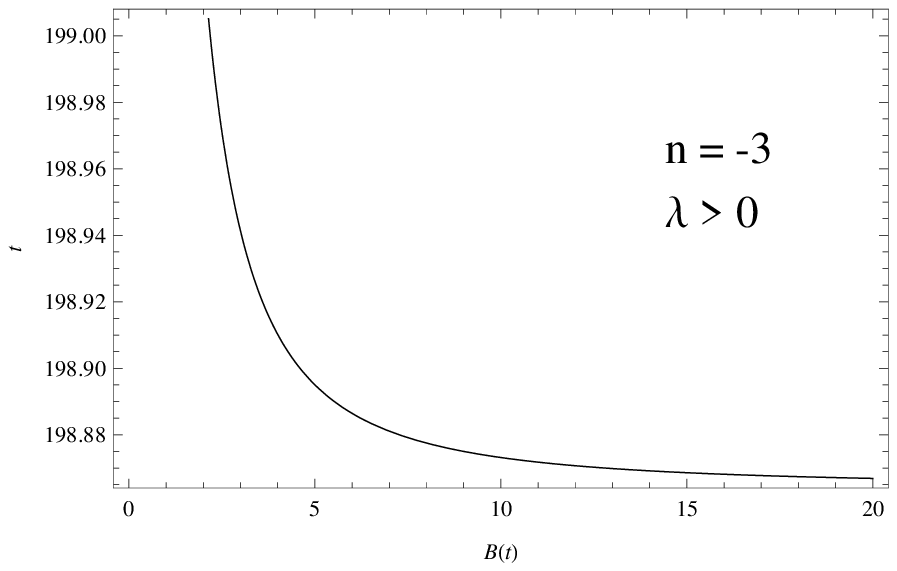}
\includegraphics[width=0.4\textwidth]{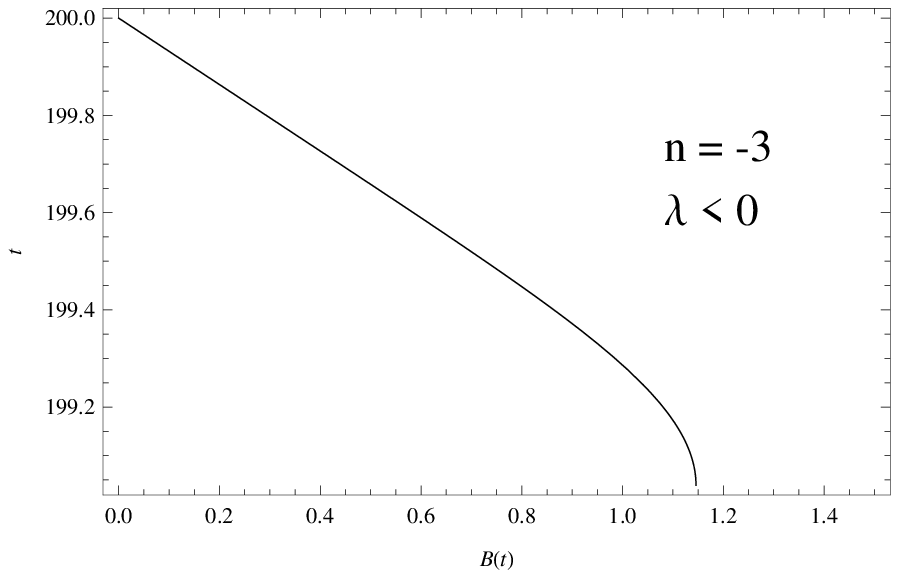}
\caption{Time evolution for different signatures of initial condition ($\lambda$) and $f(R)=-\frac{1}{2R^2}$.}
\end{center}
\label{fig:ss34}
\end{figure}

For $-\frac{1}{2}<n<0$, (see figure $(3)$) the collapsing scenario as given in Figure $(3)$ is quite different from the ones in the previous section. For a positive $\lambda$, where the sphere contracts at a steady manner, but at a particular time the radius suddenly hurries towards zero. For $\lambda<0$, the sphere collapses steadily to a certain volume, and apears to equilibriate itself at a finite volume.    \\

Figures $(4)$ and $(5)$ depict the scenario for positive values of $n$ in the allowed domain. The models show unhindered collapse, the radius goes to zero at a finite future. We have chosen two values of $n$ as examples, $n=2$ (Figure $(4)$) and $n=1$ (Figure $(5)$) for example. Both cases show that for a positive $\lambda$, the process of collapse slows down towards the end, whereas for a negative $\lambda$ the radius shrinks zero rather rapidly. For other positive values of $n$, the scenario is exactly similar qualitatively. We have included only examples.

\begin{figure}[h]
\begin{center}
\includegraphics[width=0.4\textwidth]{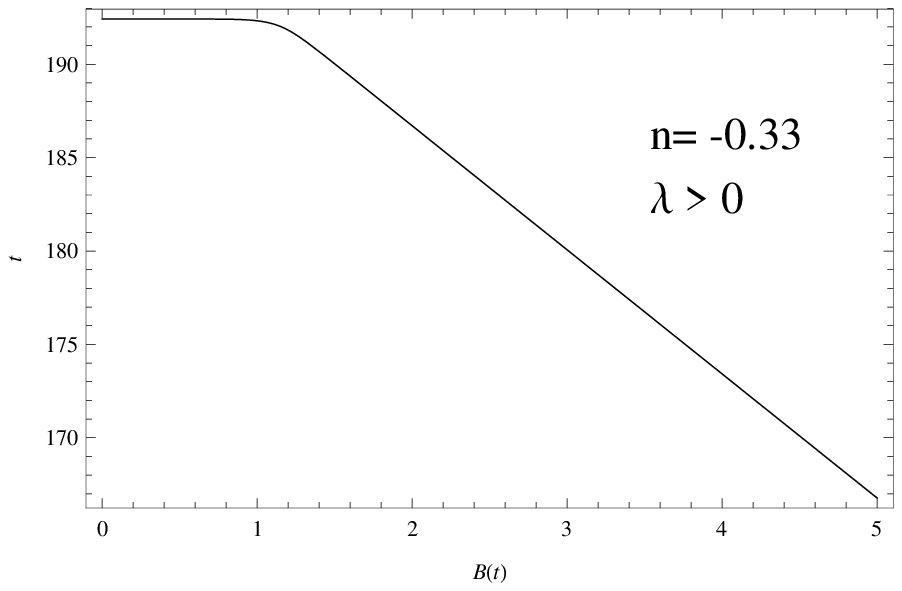}
\includegraphics[width=0.4\textwidth]{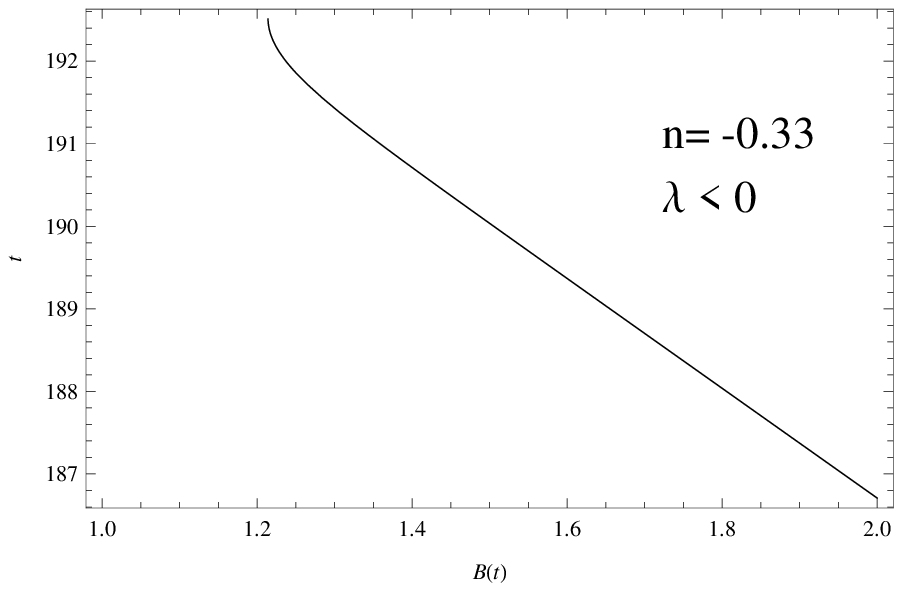}
\caption{Time evolution for different signatures of initial condition ($\lambda$) and $f(R)=\frac{3R^{2/3}}{2}$.}
\end{center}
\label{fig:ss78}
\end{figure}

\begin{figure}[h]
\begin{center}
\includegraphics[width=0.4\textwidth]{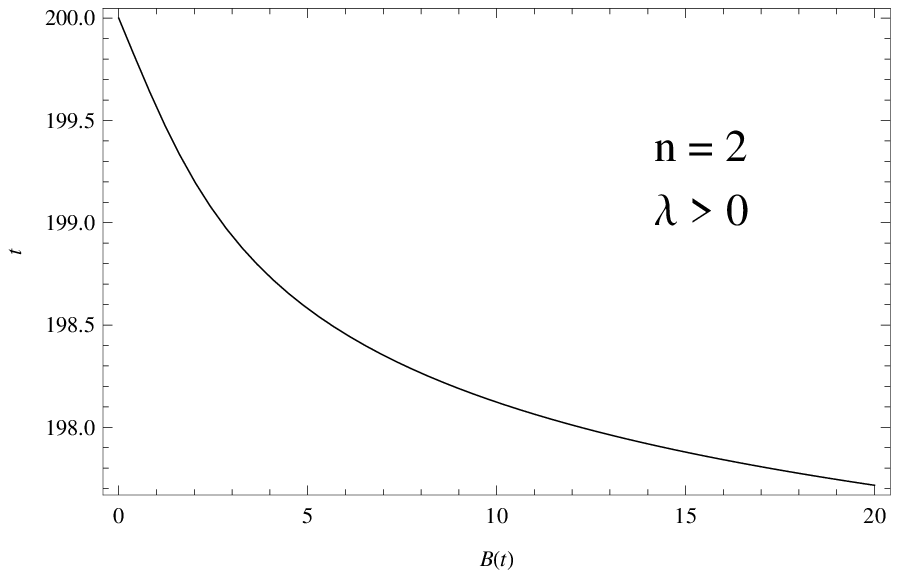}
\includegraphics[width=0.4\textwidth]{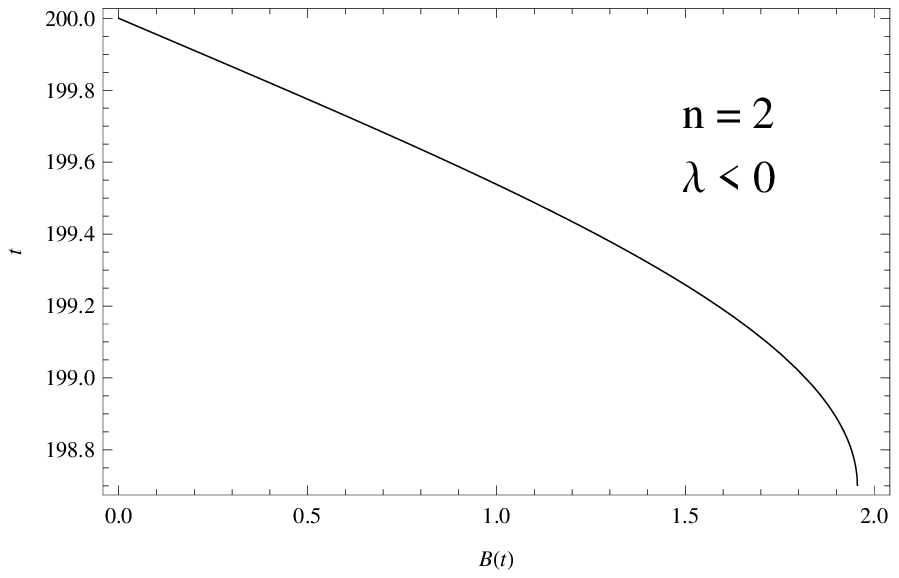}
\caption{Time evolution for different signatures of initial condition ($\lambda$) and $f(R)=\frac{R^3}{3}$.}
\end{center}
\label{fig:ss910}
\end{figure}

\begin{figure}[h]
\begin{center}
\includegraphics[width=0.4\textwidth]{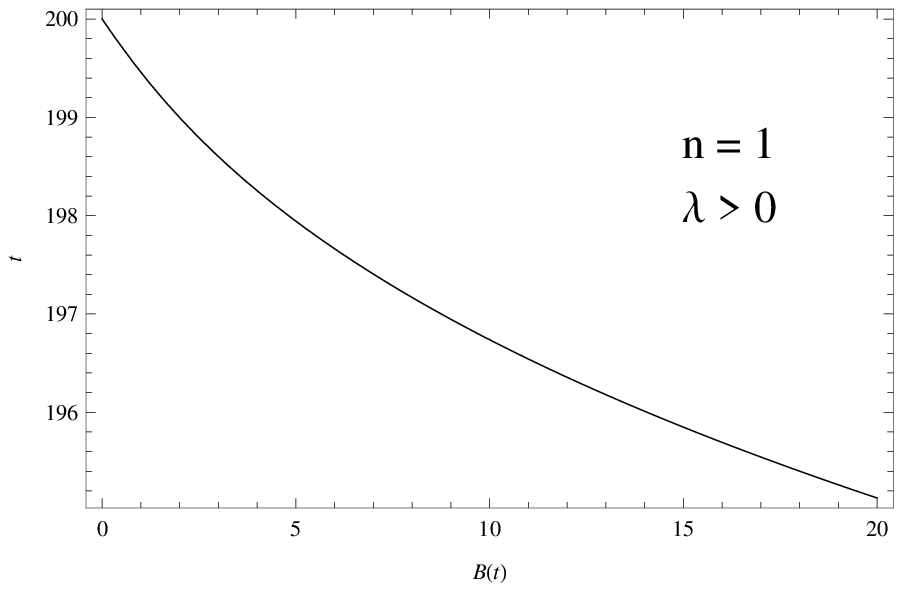}
\includegraphics[width=0.4\textwidth]{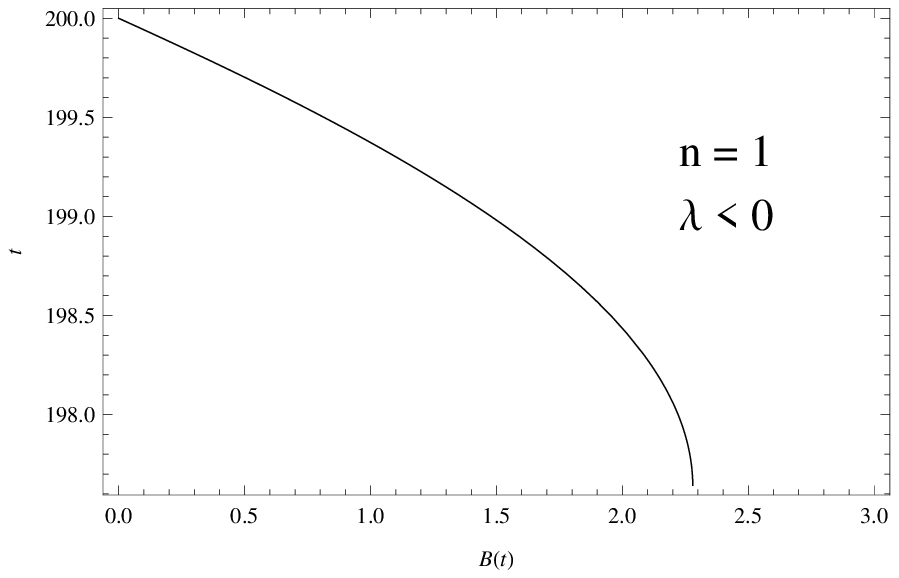}
\caption{Time evolution for different signatures of initial condition ($\lambda$) and $f(R)=\frac{R^2}{2}$.}
\end{center}
\label{fig:ss1112}
\end{figure}

From the metric (\ref{metric}), we calculate the Kretschmann Curvature Scalar as:
\begin{equation}
{\textit{K}}={R}_{abcd}{R}^{abcd}=\frac{1}{y^4}\Bigg[\frac{3}{B^4}-4\frac{\dot{B}^2}{B^4}+12\frac{\dot{B}^4}{B^4}-8\frac{\ddot{B}}{B^3}+12\frac{\ddot{B}^2}{B^2}\Bigg].
\end{equation}
Though an explicit expression of $B(t)$ as a function of time could not be written, one can assess the nature of Kretschmann Scalar from the first integral (\ref{1stint}).
\begin{equation}\label{Kret}
K=\frac{1}{y^4}\Bigg[\frac{3\alpha^2-8n\alpha(n+1)+48n^2(n+1)^2}{\alpha^2B^4}+\frac{48n\lambda(n+1)-4\alpha^2\lambda-4\alpha\lambda}{B^{4-\alpha}}+\frac{3\lambda^2\alpha^2+12\lambda^2}{B^{4-2\alpha}}\Bigg].
\end{equation}

It is clearly seen that for $B \rightarrow 0$, at least the first term in the expression for $K$ will always blow up, indicating that the singularity that one obtains is indeed a curvature singularity in nature.    \\

From the equations (\ref{density}) and (\ref{press1}), one can write the density and pressure respectively as

\begin{equation} \label{exp-density}
\rho=\frac{1}{\kappa} \Bigg[\frac{1}{y(r)^2}\Bigg(3+\frac{n}{m}\Bigg)\Bigg(\lambda B^{\alpha}+\frac{2n(n+1)}{\alpha}\Bigg)\frac{B^{\frac{n}{m}-2}}{\delta(r)^{\frac{n}{m}}}+\frac{1}{2}\frac{n}{(n+1)}\Bigg(\frac{B}{\delta(r)}\Bigg)^{\frac{n+1}{m}}+\frac{n(2n-1)}{y(r)^2}\frac{B^{\frac{n}{m}-2}}{\delta(r)^{\frac{n}{m}}}\Bigg],
\end{equation}

and, 

\begin{eqnarray}\nonumber \label{exp-press}
&& p=\frac{1}{\kappa} \Bigg[\frac{1}{y(r)^2}{\Bigg(\frac{n}{m}+\frac{n^2}{m^2}-1\Bigg)}\Bigg(\lambda B^{\alpha}+\frac{2n(n+1)}{\alpha}\Bigg)\frac{B^{\frac{n}{m}-2}}{\delta(r)^{\frac{n}{m}}}+\frac{n(2n-1)+\frac{1}{2}}{y(r)^2}\frac{B^{\frac{n}{m}-2}}{\delta(r)^{\frac{n}{m}}} \\ &&-\frac{1}{2}\frac{n}{(n+1)}\Bigg(\frac{B}{\delta(r)}\Bigg)^{\frac{n+1}{m}}-\frac{n \alpha \lambda}{m y(r)^2}\frac{B^{\alpha+\frac{n}{m}-2}}{\delta(r)^{\frac{n}{m}}}\Bigg],
\end{eqnarray}

where, $m=\frac{n(2n+1)}{2(2n-1)}$ and $\alpha=4\frac{(2n-1)}{(2n+1)}$. As these physical quantities, are all functions of $r$, the fluid distribution is not spatially homogeneous. The collapse is thus different from the Oppenheimer-Snyder collapse in general relativity\cite{opp}.    \\

For both $\rho$ and $p$, the powers involved for $B(t)$ are $(\frac{n}{m}-2)$, $(\alpha+\frac{n}{m}-2)$ and $(\frac{n+1}{m})$. For all $n>-\frac{1}{2}$, at least one of them is negative making both pressure and density proportional to inverse powers of $B(t)$ in these cases, and they do diverge when a zero proper volume is reached. In the examples discussed, for $n=1$ and $n=2$, $\frac{n}{m} -2$ is negative, and for $n=-0.33$, two of them, $(\frac{n}{m}-2)$ and $(\alpha+\frac{n}{m}-2)$ are negative. This is indeed the expected behaviour of density at the singularity.    \\
However, for $n< -\frac{1}{2}$, all these powers are positive, which leads to the peculiar result that the density and pressure both vanish at $B(t)=0$, i.e., at the singularity of zero proper volume. This is an intriguing result. If the matter sector has a conservation for itself, it really cannot happen. But since $f(R)$ gravity theories are nonminimally coupled theories, this indicates that there must be an exchange of energy between the matter and the curvature sector, and the latter drains the energy from the former during collapse, and diverges at the singularity. This is evident from the expression for the Kretschmann scalar in equation (\ref{Kret}), which diverges anyway regardless of the choice of $n$.     \\

The condition for the formation of an apparent horizon is given by
\begin{equation}
\label{app-hor}
g^{\mu\nu}R,_{\mu}R,_{\nu}=0,
\end{equation}
where $R$ is the proper radius of the two-sphere given by $y(r)B(t)$ in this case. For the metric in the present case, this yields a simple result as 
\begin{equation}
\label{app-hor-const}
\dot{B}^2=\frac{1}{2}.
\end{equation}

With the help of the equation (\ref{1stint}), this condition yields a constant $B$ corresponding to the apparent horizon in terms of $n$ as 

\begin{equation} \label{app-hor_n}
B_{ap}=\Bigg[-\frac{1}{4\lambda}\Bigg(\frac{(2n^3+3n^2-n+1)}{(2n-1)}\Bigg)\Bigg]^{\frac{(2n+1))}{4(2n-1)}}.
\end{equation}

Amongst the examples worked out in the present work, the value of $B_{ap}$ from equation (\ref{app-hor-const}) comes out to be $(-\frac{5}{4\lambda})^{\frac{3}{4}}$ and $(-\frac{9}{4\lambda})^{\frac{5}{12}}$ for the last two cases respectively. Figures (5) and (4) indicates that only for a negative $\lambda$ one has a finite time curvature singularity. So the singularities are well covered by a horizon. For a positive $\lambda$, there is no respectable horizon, but there is no requirement of that either as the volume goes to zero only asymptotically. In a similar way, in all the three other examples, we have a horizon when it is relevant. In the present model, in all the cases it occurs, the singularity is independent of $r$, and thus is not a central singularity. This kind of singularity is covered by a horizon\cite{naresh}. So the existence of the apparent horizon in all the relevant cases is quite a consistent result.

\section{Matching of the collapsing fluid with a Schwarzschild exterior}
The arbitrary function $\delta(r)$ in (\ref{BandR}) can be estimated from suitable matching of the solutions for the collapsing fluid with that of a vacuum exterior geometry at the boundary. Generally, in collapsing models, the interior is matched at a boundary hypersurface with a vacuum Schwarzschild exterior, which requires the continuity of both the metric and the extrinsic curvature on the boundary hypersurface (\cite{Darmo, Israel}). However, in f(R)-theories of gravity, continuity of the Ricci scalar across the boundary surface and continuity of its normal derivative are also required (\cite{Clifton, Deru, Seno}). The rationale behind matching with a Schwarzchild is discussed in reference \cite{rituparno1}. We also refer to the work of Ganguly {\it et al}\cite{gang}. Schwarzchild solution is given by

\begin{equation}
\label{schwarz}
 ds^{2} = (1-\frac{2M}{r})dt^{2}-(1-\frac{2M}{r})dr^{2}-r^{2}(d{\theta}^{2}+\sin^{2}\theta d{\phi}^{2}),
\end{equation}
where $M$ is the total mass contained by the interior. For an interior metric given by 

\begin{equation}
\label{met-int}
 ds^2=A^2(t,r)dt^{2}-N^2(t,r)dr^{2}-C^2(t,r)(d\theta^{2}+\sin^2\theta d\phi^{2}),
\end{equation}
the matching with the metric (\ref{schwarz}) yields (using the matching of the second fundamental form or the extrinsic curvature)
\begin{eqnarray}\nonumber \label{extrinscurv}
&&2\left(\frac{\dot{C'}}{C}-\frac{\dot{C}A'}{CA}-\frac{\dot{N}C'}{NC}\right)
=^{\Sigma}
-\frac{N}{A}\left[\frac{2\ddot{C}}{C}-\left(\frac{2\dot{A}}{A}-
\frac{\dot{C}}{C}\right) \frac{\dot{C}}{C}\right]\\ &&
+\frac{A}{N}\left[\left(\frac{2A'}{A}+\frac{C'}{C}\right)\frac{C'}{C}
-\left(\frac{N}{C}\right)^2\right],
\end{eqnarray}
where $\Sigma$ is the boundary.

If $A, N, C$ of the metric (\ref{met-int}) are replaced by the solutions obtained in the present work, one obtains
\begin{equation}
\frac{\ddot{B}}{B}+\frac{\dot{B}^2}{B^2}-\frac{1}{2B^2}-\frac{\sqrt{2}\dot{B}}{B^2}=^{\Sigma}0.
\end{equation}
The arbitrary constant $\lambda$ from (\ref{1stint}) can be estimated from this matching condition with the help of an explicit form of $B(t)$ as function of $t$.
The Misner and Sharp mass function\cite{misner}, defined as 
\begin{equation}
\label{MS-mf}
m(t,r)=\frac{C}{2}(1+g^{\mu\nu}C_{,\mu}C_{,\nu})=\frac{C}{2}\left(1+\frac{\dot{C}^2}{A^2}
-\frac{C'^2}{N^2}\right),
\end{equation}
yields the mass contained by the surface defined by the radial coordinate $r$ in the present case as
\begin{equation}
\label{MS-mf2}
m(t,r)=\frac{yB}{2}(\frac{1}{2}+\dot{B}^2).
\end{equation}
The Scwarzschild mass $M$, i.e., the total mass contaiined by the collapsing fluid is given by the right hand side of equation (\ref{MS-mf2}) calculated at the boundary.   \\

The matching of Ricci Scalar and its normal derivative was studied in detail by Deruelle, Sasaki and Sendouda (\cite{Deru}). They generalized the Israel junction conditions (\cite{Israel}) for this class of theories by direct integration of the field equations. It was significantly utilised by Clifton et. al. (\cite{Clifton}) and Goswami et. al. (\cite{rituparno2}) quite recently. \\

The continuity requirements yield
\begin{equation}
R=R'=^{\Sigma}0.
\end{equation}

From the $G_{01}$ equation we already have
\begin{equation}\label{BandR}
B=\delta(r)R^m.
\end{equation}
Since the Ricci scalar is separable one can see that it should rather be of the form
\begin{equation}\label{riccimatch}
R = (r_{\Sigma}^2-r^2)^2 g(r) T(t),
\end{equation}
where $g(r)$ is a well-defined function of radial cordinate $r$ and $T(t)$ describes the time evolution of the scalar. The relevance of chosing this type of functional form for a spherical star in $f(R)$ gravity was elaborated in \cite{rituparno2}. From (\ref{BandR}) and (\ref{riccimatch}), we can write
\begin{equation}
\delta(r)=\Bigg(r_{\Sigma}^2-r^2\Bigg)^{\frac{n(2n+1)}{(1-2n)}}g(r)^{\frac{n(2n+1)}{2(1-2n)}},
\end{equation}
which describes the radial profile of the spherical collapsing body.

\section{Discussion}
A spherical collapse in $f(R)$ gravity is investigated analytically here. In fact, with the assumption of the existence of a homothetic Killing vector,  exact solutions for the collapsing models are obtained for $f(R)\sim R^{n}$. The solutions are not readily invertible in the form of the proper radius as a function of time. So the approach to the singularity of a zero proper volume is investigated with the help of numerical plots of time against proper radius. Some of the collapsing modes indeed succumb to the singularity at a finite future. In some cases, where $f(R)$ varies as positive powers of $R$, the singularity is reached only at an infinite future. This variation certainly depends on the initial condition, determined by $\lambda$. Where $f(R)$ varies as negative powers of $R$, the collapsing object might equilibriate at a constant proper radius and never hit the singularity. Singularities formed at a finite future are covered by an apparent horizon. \\

One intriguing result is that in some cases, the density and pressure of the fluid present vanish at the singularity. It appears that the energy is drained in by the curvature component in the stress-energy...

\end{document}